\documentclass[hidelinks]{article}

\usepackage{arxiv}

\usepackage[utf8]{inputenc} 
\usepackage[T1]{fontenc}    
\usepackage{hyperref}       
\usepackage{url}            
\usepackage{booktabs}       
\usepackage{amsfonts}       
\usepackage{nicefrac}       
\usepackage{microtype}      
\usepackage{lipsum}		
\usepackage{graphicx}
\usepackage{doi}
\usepackage{subfigure}
\usepackage{caption}

\title{Ciphertext Policy Attribute Based Encryption with Intel SGX}

\author{\href{https://orcid.org/0009-0009-8972-5609}{\includegraphics[scale=0.06]{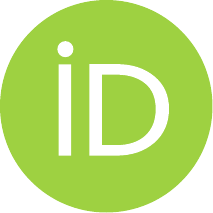}\hspace{1mm}Vivek Suryawanshi}
\\
Department of Computer Science and Engineering\\
	Indian Institute of Technology \\
	Kharagpur, 721302 \\
\texttt{viveksurya552@gmail.com} \\
\And
\href{https://orcid.org/0000-0002-4315-7329}{\includegraphics[scale=0.06]{orcid.pdf}\hspace{1mm}
 Shamik Sural} \\
	Department of Computer Science and Engineering\\
	Indian Institute of Technology \\
	Kharagpur, 721302 \\
	\texttt{shamik@cse.iitkgp.ac.in} \\
}

\date{}

\renewcommand{\headeright}{}
\renewcommand{\undertitle}{}
\renewcommand{\shorttitle}{Ciphertext Policy Attribute Based Encryption with Intel SGX}

\begin{document}
\maketitle

\begin{abstract}
Modern computing environments demand robust security measures to protect sensitive data and resources. Ciphertext-Policy Attribute-Based Encryption (CP-ABE) is a well-established encryption technique known for its fine-grained access control capabilities. However, as the digital landscape evolves, there is a growing need to enhance the security of CP-ABE operations. We propose an approach that utilizes CP-ABE with Intel SGX. It allows data to be encrypted and decrypted securely within the SGX enclave based on the rules in policy by ensuring that only authorized users gain access. We evaluate its performance through different experiments by focusing on key parameters such as the number of rules, attributes and file size. Our results demonstrate the performance and scalability of integrating SGX with CP-ABE in enhancing data security with only minimal increase in execution time due to enclave overhead.
\end{abstract}


\section{Introduction}
In today's digital era, safely keeping and accessing sensitive information is a top priority for software applications. Sensitive information like passwords and digital keys needs to be protected for ensuring privacy and security. Access control is a security technique that ensures only authorized users are granted access while unauthorized ones are denied permission. Attribute-Based Encryption (ABE) introduces a dynamic model in cryptography by offering a convenient way to data security and access control. Unlike traditional encryption techniques, ABE enables data owners to associate attributes with encrypted data by granting access based on these attributes rather than fixed identities. It is similar to giving access to particular information only to those who meet specific criteria. This flexible approach helps in sharing sensitive data securely. ABE is categorized into two types – Key Policy ABE (KP-ABE) and Ciphertext Policy ABE (CP-ABE).
KP-ABE encrypts data with attributes and users are issued keys that match specific policies defined by the data owner \cite{kpabe-paper}. This approach allows data owners to enforce finely tuned access policies by ensuring that only users with keys matching the predefined policies can decrypt and access the data. KP-ABE proves particularly useful in hierarchical organizations where different levels of access are required. However, managing keys for complex policies and attribute assignments can present a challenge. CP-ABE takes a different approach by encrypting data with access policies defined using attributes while encryption and decryption permission is granted based on an evaluation of this policies against attributes of user \cite{cpabe-paper}. This innovative approach empowers data owners to share sensitive information with specific users based on a set of attribute criteria. CP-ABE is especially useful in scenarios where data owners want to share information with a specific audience that fulfills predetermined requirements. ABE plays a crucial role in modern data security, particularly in environments where access control and confidentiality are paramount. 

Hardware-level enclaves are trusted execution environments in modern processors such as Intel SGX, ARM TrustZone and AMD SEV. These enclaves provide isolated and encrypted regions of memory for secure computations in applications. Enclaves are implemented in hardware and provide a high level of security as they are isolated from the operating system and other processes, and their memory contents are encrypted. So enclaves are resistant to attacks such as memory tampering or unauthorized access. Enclaves are increasingly being used in various applications. Intel Software Guard Extensions (SGX) is a security technology embedded in Intel CPUs which eases the creation of isolated enclaves for sensitive computations by protecting data from privileged software and operating systems \cite{intel-sgx-dev-guide}. It provides a robust foundation for building trusted computing environments by enhancing the security of sensitive computations and data in a wide range of applications. The combination of CP-ABE policy enforcement with Intel SGX provides a compelling solution to address the challenges of securing data access and protection in today's digital landscape. There is limited work done so far in combining CP-ABE with SGX. The work of \cite{10.1007/978-3-030-92708-0_4} explores the possibility of using blockchains with ABE and SGX. By leveraging SGX's capabilities, we establish dedicated enclaves for CP-ABE policy enforcement by ensuring that access control and policies are shielded from potential attacks or unauthorized access.

\section{Preliminaries}
In this section, we first briefly describe the working of CP-ABE, followed by Intel SGX.

\subsection{Ciphertext Policy Attribute Based Encryption}
Ciphertext Policy Attribute Based Encryption (CP-ABE) introduces a revolutionary model in data security where protection of information is intricately linked to access policies defined using attributes. In this innovative encryption framework, data owners use the power to formulate specific access policies for granting decryption privileges exclusively to users possessing attributes that matches with the rules in the defined policies. One of the notable strengths of CP-ABE lies in its provision of fine-grained access control. 

The functional flow of CP-ABE has three distinct stages : setup, encryption and decryption \cite{cpabe-kit}. Each stage plays a pivotal role in the secure processing of data. During the setup phase, two critical components are generated – the public key (pub key) and the master key. These components are built using the Pairing-Based Cryptography (PBC) \cite{pbc-download} library and required for subsequent encryption and decryption operations. The public key becomes an integral element for both encryption and decryption processes while the master key is vital for enabling the decryption of encrypted content.
The encryption phase of CP-ABE involves the creation of a policy tree and an encryption key. The policy tree consists of rules that describe the access requirements for decrypting the data. For instance, a rule could be expressed as "designation:professor  department:cs  file-type:pdf  3of3". Here “3of3” indicates that all three attributes are essential for satisfying the rule. An encryption key is generated using this policy tree and the public key. This key is then applied to encrypt the plaintext of a file, resulting in a cipher text. This cipher text is securely stored in a newly encrypted file by ensuring the confidentiality and privacy of the underlying data.
Subsequently, in the decryption phase, a private key is generated using the master key and user provided attributes. Policy tree from the encryption phase is used to evaluate whether the user's attributes meet the access policy criteria using private key. Upon successful evaluation, a decryption key is generated by combining the policy tree and the pub key. This decryption key is used to decrypt the cipher text stored in the encrypted file to uncover the original plain text. The decrypted plain text content is then saved into a new decrypted file for allowing authorized users to access the original plaintext data. The secure decryption process ensures that only users with the necessary attributes are authorized to access and comprehend the data. This decryption flow showcases CP-ABE as a dynamic tool for safeguarding sensitive information by providing a granular and adaptable approach to data access control and security.

\subsection{Intel SGX}
Intel Software Guard Extensions (SGX) is a cutting-edge security feature available in Intel CPUs starting from the 6th generation and newer. Intel SGX is set of extensions to Intel's instruction set architecture (ISA) that enables software developers to create enclaves. These enclaves are secure, isolated and encrypted regions in memory where sensitive computations can be executed with utmost security. Enclaves are designed to be completely isolated from the rest of the system and ensures that data within an enclave are protected from external tampering or unauthorized access. 

Application design within the Intel SGX ecosystem involves two primary components: the trusted component and the untrusted component \cite{intel-sgx-tutorial}. The Trusted Component is also referred as the enclave where the secrets of an application are stored. It is crucial to keep this enclave as small as possible to enhance security. On the other hand, the Untrusted Component contains the remaining parts of the program including the operating system and other system software, considered untrusted from the enclave's perspective. The execution flow of an Intel SGX application has several steps. During the application build, the code is divided into trusted and untrusted components. Enclaves communicate with untrusted code through Enclave Calls (ECALLs) and Outside Calls (OCALLs). ECALLs allow the application to invoke a predefined function inside the enclave while OCALLs enable calls from inside the enclave to the external application \cite{sgx:2016/086}.

In Intel SGX, attestation ensures enclave integrity and certify secure execution. Attestations in Intel SGX are of two types - Local attestation, which is mainly used for verifying enclaves within the same computing platform, while Remote attestation is used for validating enclaves across different platforms or hosts. Data and code of enclaves are not considered as secret. Secrets are introduced later and can be encrypted for external storage using a sealing mechanism. Intel SGX sealing feature is useful to make sensitive data tamper resistant even when stored outside of enclave. Intel SGX provides a powerful security tool for protecting sensitive computations such as cryptographic operations, key management, and secure data processing. It allows software developers to build applications that can securely process sensitive data without exposing them to potential security threats.

\section{CP-ABE with Intel SGX}
In this section, we first describe our proposed method for safeguarding CP-ABE components and policy evaluation using Intel SGX, and then provide a description of the web based file encryption and decryption tool using CP-ABE.

\begin{figure}[t]
\centerline{\includegraphics[width=0.85\linewidth]{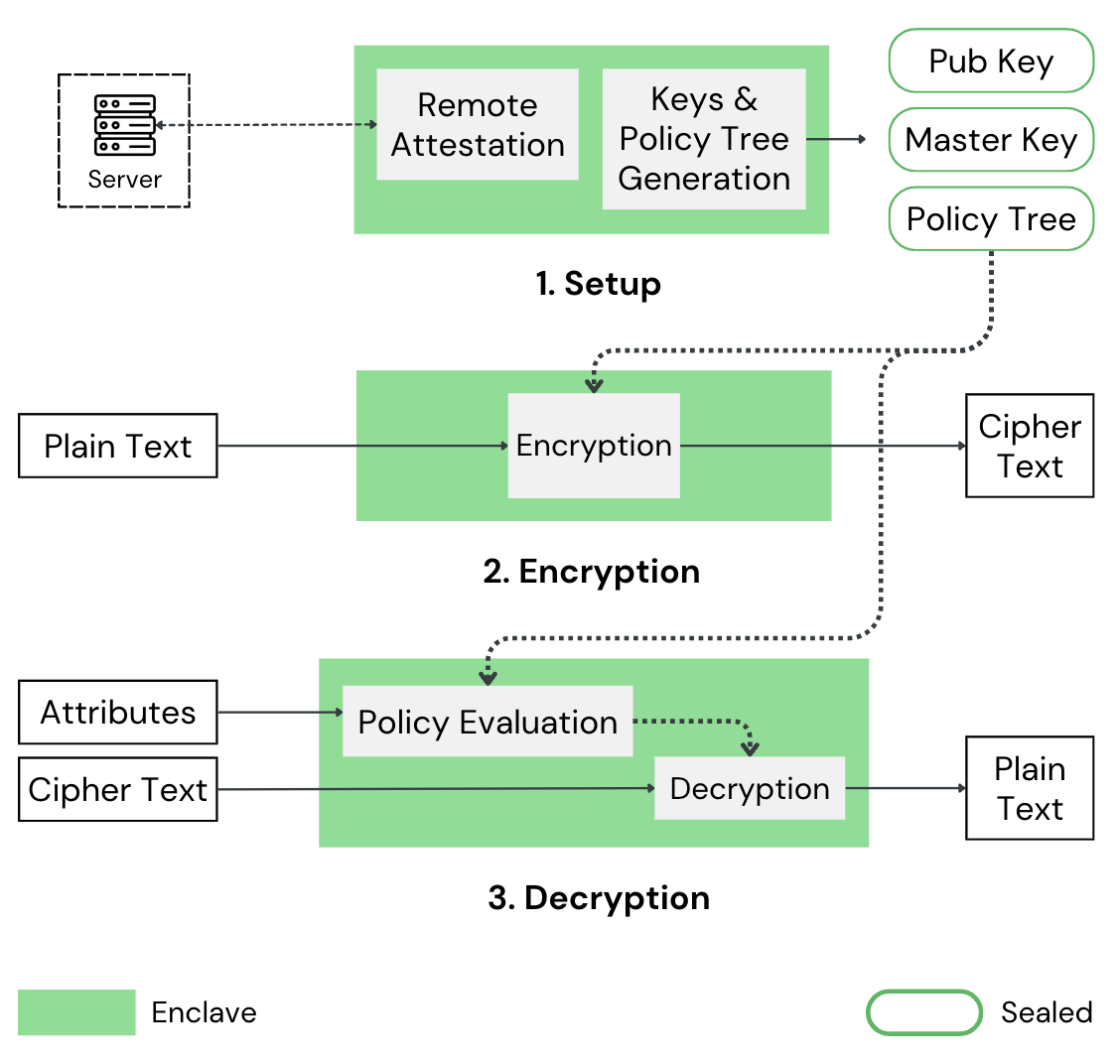}}
\caption{\fontsize{9}{11}\selectfont Flow Diagram of CP-ABE with Intel SGX}
\label{fig:flow-cpabe-sgx}
\end{figure}

\subsection{Proposed Method}
In the realm of secure data access and protection, CP-ABE stands out for its fine-grained access control capabilities. To address evolving security challenges, we propose a method that integrates the robust security features of Intel SGX with CP-ABE. This combination creates a execution environment within SGX by shielding encryption and decryption functions from threats and unauthorized access thereby enhancing CP-ABE security. The core functionality of CP-ABE is typically written in C and it also relies on external libraries such as OpenSSL, Pairing Based Cryptography (PBC) and the GNU Multiple Precision Arithmetic Library (GMP). But challenges are introduced while porting such functionality within an SGX enclave ecosystem due to external library dependency. We address this issue by installing SGX-compatible libraries and ensure seamless and secure CP-ABE core functionality within SGX enclaves \cite{pbc-sgx}.

\begin{figure}[t]
\centerline{\includegraphics[width=0.80\linewidth]{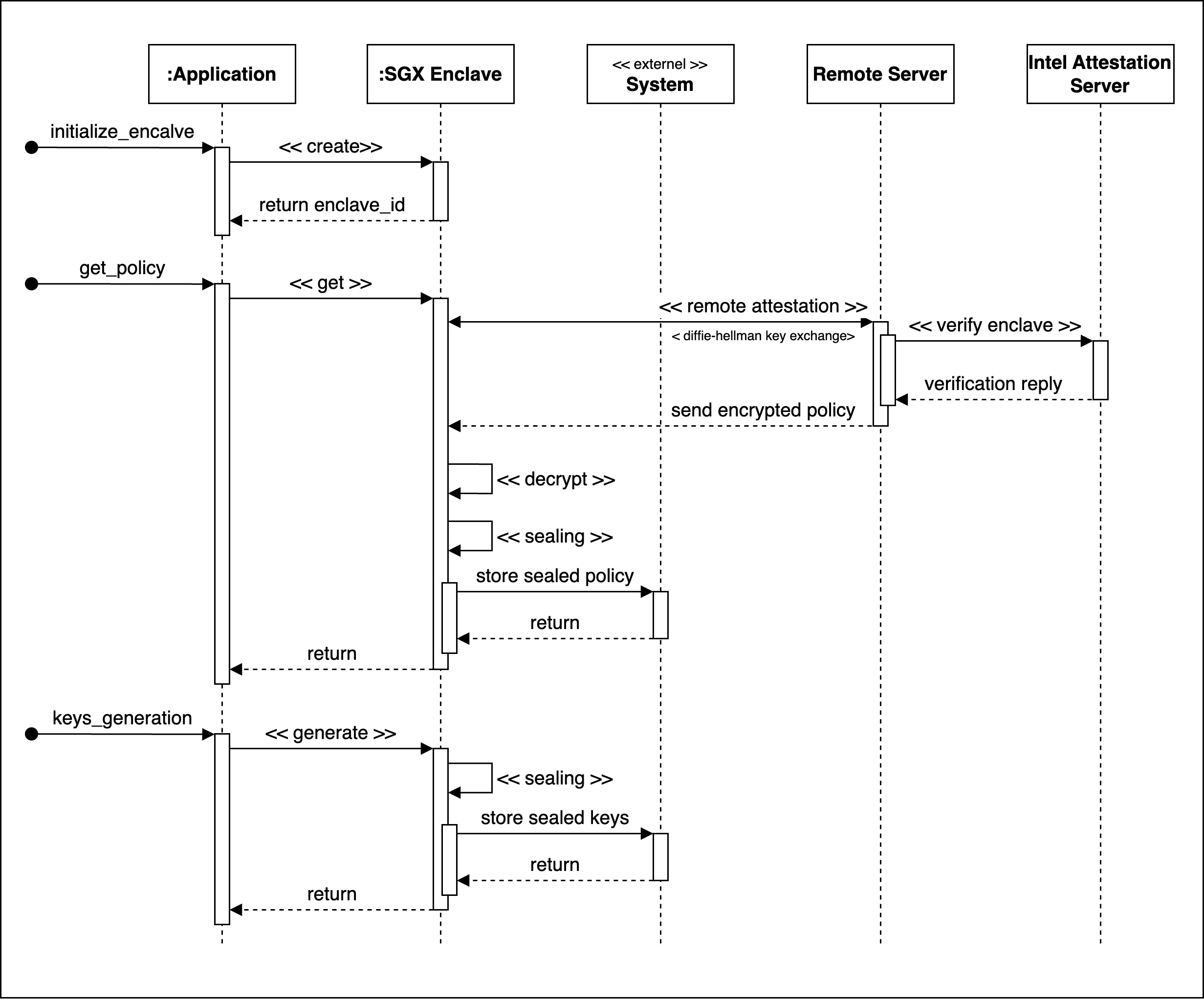}}
\caption{\fontsize{9}{11} Sequence Diagram for Setup Phase}
\label{fig:seq-setup}
\end{figure}

The integration of CP-ABE with Intel SGX follows a well-defined flow to enhance the security of encryption and decryption processes (Figure \ref{fig:flow-cpabe-sgx}). The initial setup involves creating a secure SGX enclave where essential cryptographic keys - public key (pub key) and master key are generated (Figure \ref{fig:seq-setup}). To boost the security of our proposed method, we implement remote attestation which involves a secure exchange of attestation evidence between the enclave and a remote server ensuring that the enclave's identity is validated before any secure operation takes place. Also, a policy defining access rules is securely obtained from a remote server. Critical components such as public keys, master keys and access policies are stored outside the enclave using sealing mechanism of SGX to safeguard against potential attacks.

\begin{figure}[t]
    \centering
    \subfigure[\textbf{Sequence Diagram for Encryption Phase}]{\includegraphics[width=0.45\textwidth]{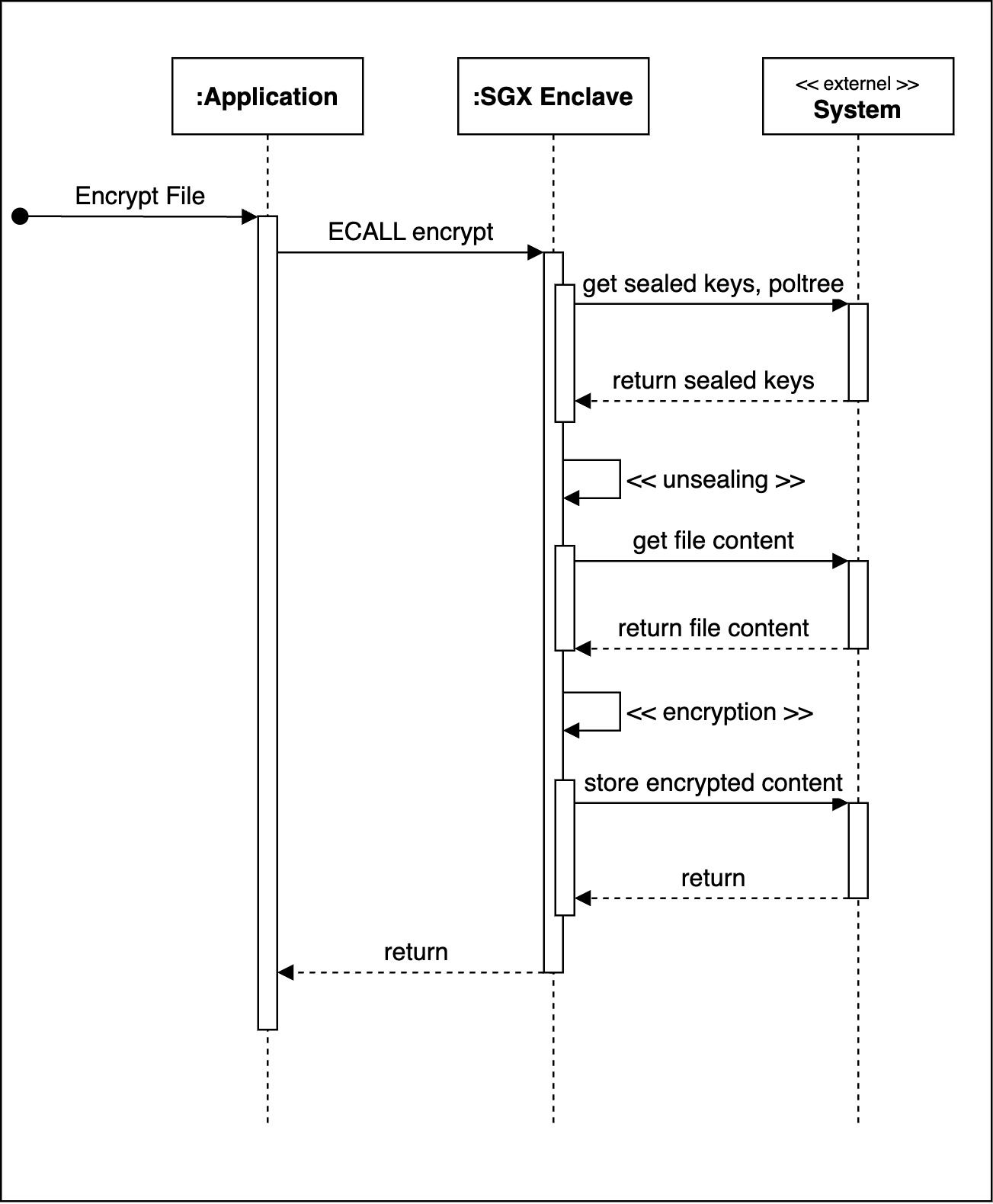}}
    \hfil
    \subfigure[\textbf{Sequence Diagram for Decryption Phase}]{\includegraphics[width=0.45\textwidth]{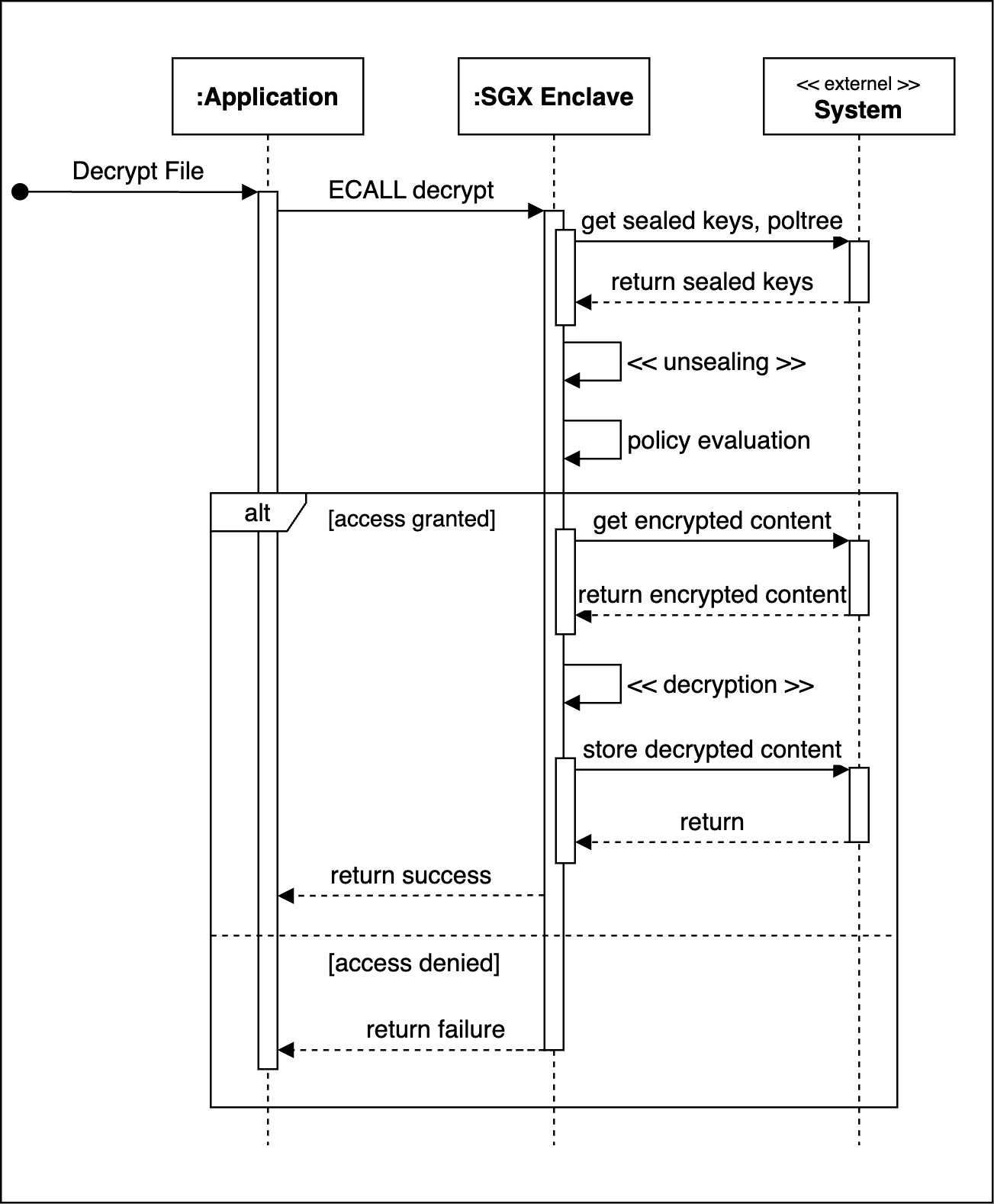}}
    \hfil
    \caption{\fontsize{9}{11} Sequence Diagrams}
    \label{fig:seq-2-diag}
\end{figure}

The execution flow incorporates ECalls for core functionality of CP-ABE within SGX. These ECalls serve as secure communication channels by allowing untrusted components outside the enclave to interact securely with the enclave's execution environment. For encryption, an ECall takes plaintext data and unseals key components and policy using sealing. It generates a policy tree along with an encryption key, and securely encrypts the data. The resulting encrypted cipher text is then sent back to untrusted code for storage (Figure \ref{fig:seq-2-diag}(a)). On the decryption side, an ECall takes user attributes along with the cipher text, and evaluates attributes against the policy to return the decrypted plaintext to untrusted code. This decrypted content is stored securely in a plain file format (Figure \ref{fig:seq-2-diag}(b)). Additionally, OCalls enable the enclave to communicate securely with the untrusted environment by ensuring that both plaintext and ciphertext are sent to untrusted components for further process. This two-way communication makes sure that the enclave can securely handle data transfers while preserving the integrity of the encryption and decryption processes. Our method enhances CP-ABE security by leveraging SGX's secure execution by resolving dependencies, implementing remote attestation and securing critical components.

\subsection{Tool Description}
\begin{figure*}[t]
    \centering
    \subfigure[\textbf{File encryption using CP-ABE with Intel SGX}]{\includegraphics[width=0.45\textwidth, trim=0.6cm 0cm 1.6cm 0cm, clip]{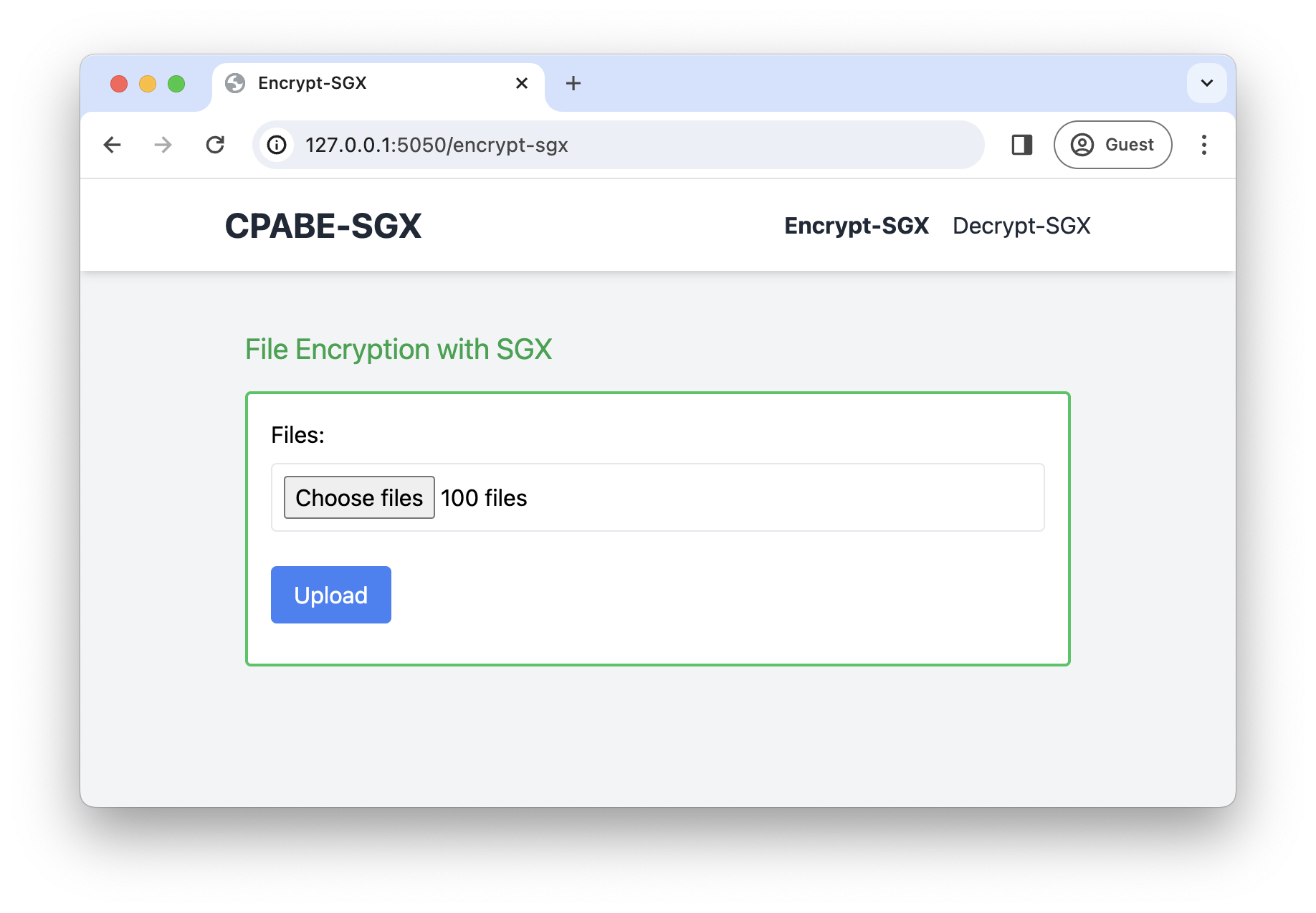}}
    \hfil
    \subfigure[\textbf{File decryption using CP-ABE with Intel SGX}]{\includegraphics[width=0.45\textwidth, trim=0.6cm 0cm 1.6cm 0cm, clip]{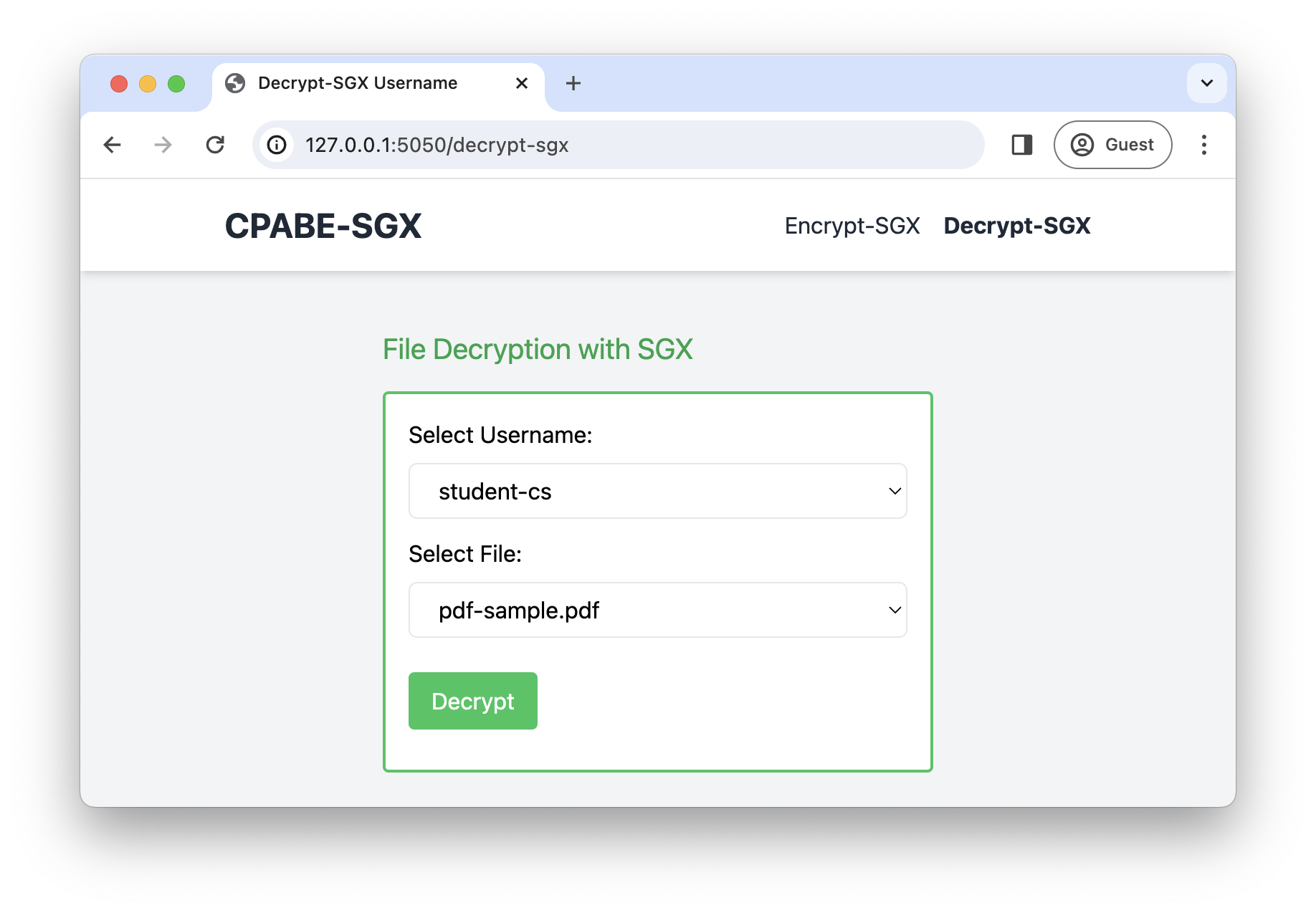}}
    \hfil
    \subfigure[\textbf{Decrypted file download}]{\includegraphics[width=0.45\textwidth, trim=0.6cm 0cm 1.6cm 0cm, clip]{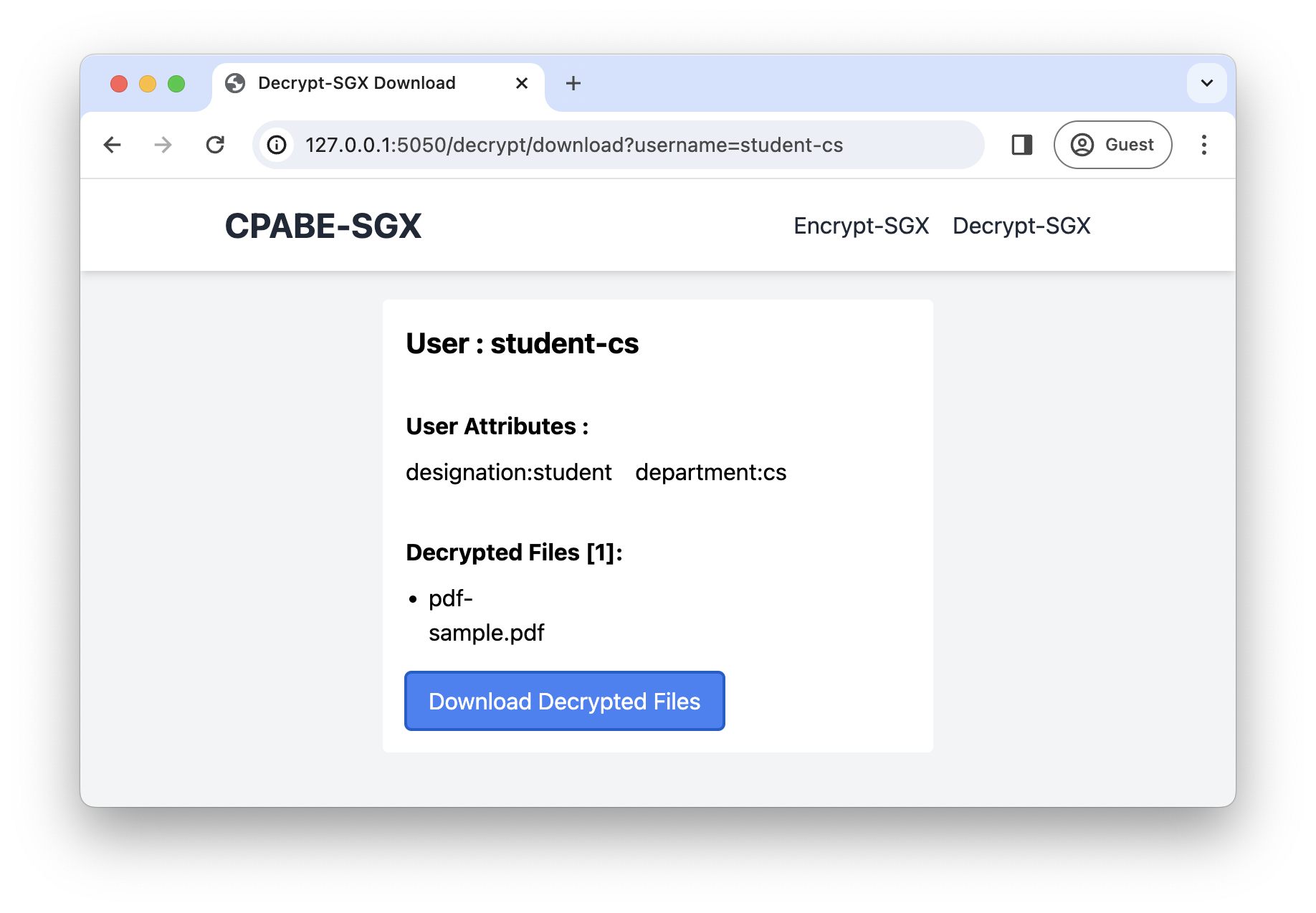}}
    \caption{\fontsize{9}{11} Screenshots of Web-based Tool}
    \label{fig:tool}
\end{figure*}
We have developed a web application that leverages CP-ABE in conjunction with Intel SGX to enhance file security. It enables users to encrypt files with customized access policies based on attributes and to decrypt the encrypted file if attributes of users are satisfied with policy rules. We engineered this web application within a Linux environment to ensure compatibility and stability. It has been developed using Python with the Flask framework by providing a user friendly interface.
The project's adaptability is achieved through the interplay of CP-ABE and SGX code which is implemented in the C programming language and seamlessly integrated using the ctypes module in Python.

Firstly, our web application initiates the setup phase which involves creating the SGX enclave and generation \& sealing of public key \& master key and conducting remote attestation to validate its integrity. During this phase, the access control policy is securely obtained from a remote server. A user wishing to encrypt a file accesses the application's '/encrypt-sgx' page and uploads the file through a form. In the background, the CP-ABE encryption function is invoked inside SGX enclave for each uploaded file. Also, Ocalls are invoked from the enclave to get the file content and to store encrypted file content. This function ensures that the file is securely encrypted within the enclave in compliance with the access control policy.  
Any other user who seeks to decrypt a file accesses the '/decrypt-sgx' page. Using the provided form, such users select their own set of attributes and choose the file they wish to decrypt. The selected attributes along with the attributes associated with the file are evaluated against the access control policy. If the criteria are met, then the file is securely decrypted and made available for download. This decryption process is executed by invoking the CP-ABE decryption function within the enclave. Also Ocalls are invoked from the enclave to get encrypted file content and to store decrypted file content.
Our web application provides a powerful solution for data security, allowing only authorized users to access and decrypt files with precision. Example screenshots of the web app are shown in Figures \ref{fig:tool}(a)-(c).


\begin{figure*}[t]
    \centering
    \subfigure[\textbf{Execution Time vs Number of Rules}]{\includegraphics[width=0.45\textwidth, trim=0.6cm 0cm 1.6cm 0cm, clip]{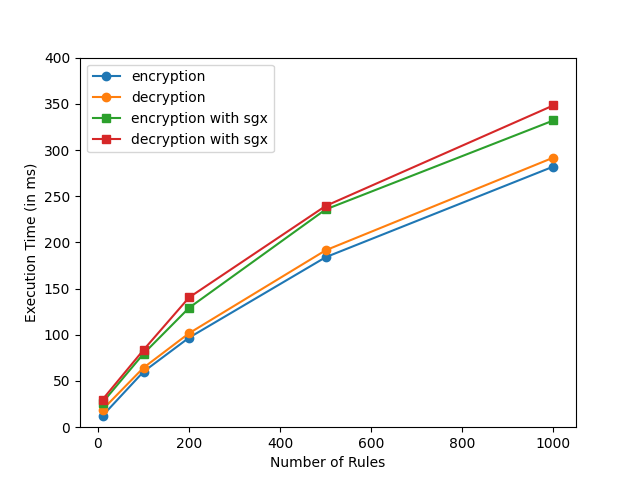}}
    \hfil
    \subfigure[\textbf{Execution Time vs Number of Attributes}]{\includegraphics[width=0.45\textwidth, trim=0.6cm 0cm 1.6cm 0cm, clip]{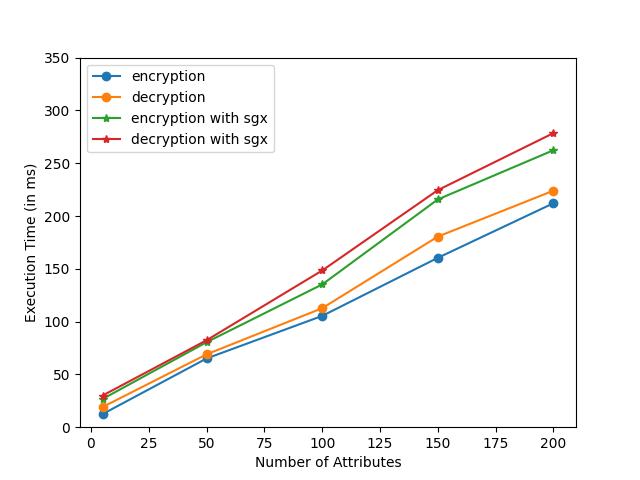}}
    \hfil
    \subfigure[\textbf{Execution Time vs File Size}]{\includegraphics[width=0.45\textwidth, trim=0.6cm 0cm 1.6cm 0cm, clip]{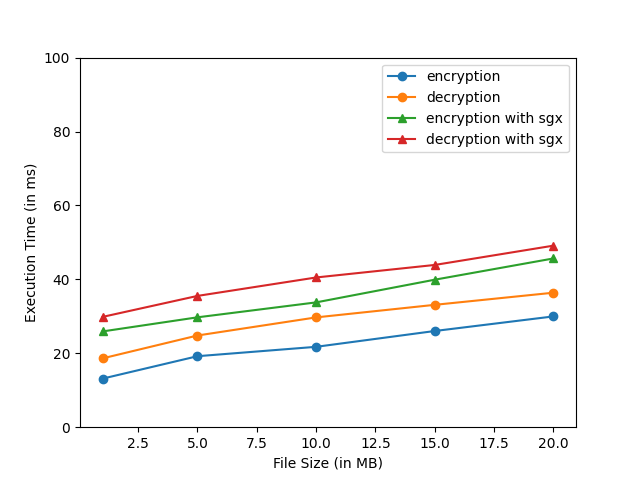}}
    \caption{\fontsize{9}{11} Results of Experiments}
    \label{fig:three_images}
\end{figure*}

\section{Experimental Results}
We conducted a set of experiments on a system equipped with an Ubuntu 20.04 operating system running on an Intel i5 8th Gen processor with integrated SGX hardware support.

Figure \ref{fig:three_images}(a) illustrates the execution time versus number of rules in the access policy. We varied the number of rules while keeping the number of attributes fixed as 5, and the file size constant at 500 KB. The results depict the execution times for both encryption and decryption processes with and without SGX. As the number of rules increases, both encryption and decryption times show a gradual rise. However, when SGX is employed, there is some increase in execution time compared to the traditional CP-ABE approach due to the additional overhead of SGX enclave operations.

Figure \ref{fig:three_images}(b) shows the relationship between execution time and the number of attributes in the access policy. We varied the number of attributes while maintaining the number of rules constant at 10 and a file size of 500KB. As the number of attributes increases, there is a linear increase in execution times for both encryption and decryption.  

Figure \ref{fig:three_images}(c) showcases the impact of file size on the execution time of encryption and decryption processes. We varied the file size from 1 MB to 50 MB for 5 attributes and 10 rules. The results indicate an incremental increase in execution times as the file size grows. Additionally, when SGX is utilized, then the execution times increase slightly due to the overhead introduced by SGX enclave operations.

\section{Conclusion and Future Directions}
In this paper, we explored the possibility of integration of Intel SGX with Cipher-Policy Attribute-Based Encryption CP-ABE to enhance the security of data access and protection mechanisms. While this integration introduced some overhead, it significantly improved data confidentiality and integrity. Our experiments showed that SGX-enabled CP-ABE is scalable across different policy, attribute, and file sizes, making it a promising solution for securing data in untrusted environments. For future research, optimizing techniques for policy evaluation will be a key focus. This involves enhancing algorithms to evaluate access policies more efficiently by resulting in faster decision-making processes. 

\bibliographystyle{unsrt}
\bibliography{CPABE_SGX}

\begin{thebibliography}{1}

\bibitem{kpabe-paper}
Amit Sahai and Brent Waters.
\newblock Fuzzy identity-based encryption.
\newblock In {\em 24th Annual International Conference on the Theory and Applications of Cryptographic Techniques, Aarhus, Denmark, May 22-26, 2005}, pages 457--473. Springer, 2005.

\bibitem{cpabe-paper}
John Bethencourt, Amit Sahai, and Brent Waters.
\newblock Ciphertext-policy attribute-based encryption.
\newblock In {\em 2007 IEEE Symposium on Security and Privacy (SP'07)}, pages 321--334. IEEE, 2007.

\bibitem{intel-sgx-dev-guide}
Intel Corporation.
\newblock Intel software guard extensions developer guide.
\newblock \url{https://download.01.org/intel-sgx/sgx-linux/linux-2.6/docs/Intel\_SGX\_Developer\_Guide.pdf}, Accessed: 2023.
\newblock Online document.

\bibitem{10.1007/978-3-030-92708-0_4}
Liang Zhang, Haibin Kan, Yang Xu, and Jinhao Ran.
\newblock Revocable data sharing methodology based on sgx and blockchain.
\newblock In {\em Network and System Security: 15th International Conference, NSS 2021, Tianjin, China, October 23, 2021, Proceedings}, page 61–78, Berlin, Heidelberg, 2021. Springer-Verlag.

\bibitem{cpabe-kit}
{John Bethencourt}.
\newblock Advanced crypto software collection.
\newblock \url{https://acsc.cs.utexas.edu/cpabe/}, Accessed: 2023.
\newblock Online Resource.

\bibitem{pbc-download}
{Ben Lynn}.
\newblock Pbc library - pairing-based cryptography - downloads.
\newblock \url{https://crypto.stanford.edu/pbc/download.html}, Accessed: 2023.
\newblock Online Resource.

\bibitem{intel-sgx-tutorial}
Intel Corporation.
\newblock Intel software guard extensions tutorial - part 1: Foundation.
\newblock \url{https://www.intel.com/content/www/us/en/developer/articles/training/intel-software-guard-extensions-tutorial-part-1-foundation.html}, Accessed: 2023.
\newblock Website.

\bibitem{sgx:2016/086}
Victor Costan and Srinivas Devadas.
\newblock Intel {SGX} explained.
\newblock Cryptology ePrint Archive, Paper 2016/086, 2016.
\newblock \url{https://eprint.iacr.org/2016/086}.

\bibitem{pbc-sgx}
{Teh Sunn Liu}.
\newblock Github - tehsunnliu/pbc-sgx: Pairing-based cryptography (pbc library) with sgx support.
\newblock \url{https://github.com/tehsunnliu/pbc-sgx}, Accessed: 2023.
\newblock Online Resource.

\end{thebibliography}

\end{document}